\documentclass[preprint,12pt]{aastex}
\usepackage{natbib} 
             \def \etal {{\it et~al.}}

             \def \lesssim{\mathrel{<\kern-1.0em\lower0.9ex\hbox{$\sim$}}}
             \def \gtrsim{\mathrel{>\kern-1.0em\lower0.9ex\hbox{$\sim$}}}

\begin{document}

\doublespace

\title{Nebular vs. Stellar Wind Abundances in NGC 6543}

\author{H. Maness} 
\affil{Department of Physics, Grinnell College, Grinnell, IA 50112.
MANESS@grinnell.edu} 

\author{S. D. Vrtilek}
\affil{Harvard-Smithsonian Center for Astrophysics, Cambridge, MA 02138.
svrtilek@cfa.harvard.edu}

.
\newpage
\section{abstract}
An early analysis of {\it Chandra} observations of the planetary nebula 
NGC 6543 suggested that 
the location and the derived temperature for the X-ray emitting
region is inconsistent with the abundances measured for this object 
\citep{Chu01}.  
We revisit {\it Chandra} observations of this source (following
significant updates to both the reprocessing and extraction software)
in order
to propose a resolution to this apparent anomaly.
Our 
re-analysis using  
abundances found from observations in the infrared, optical, and 
ultraviolet suggests that the location and temperature of the
X-ray emission from NGC 6543 is  
consistent with nebular abundances expected for the source. 

Planetary nebulae: individual (NGC6543)

\clearpage

\section{Introduction}

Over the last several years, sensitive X-ray observations of planetary
nebulae by Chandra and XMM-Newton have led to results that both
astonish and puzzle astronomers. One such result was presented by Chu 
et al. (2001) on the Chandra-observed X-ray spectrum of the Cat's Eye 
nebula, NGC 6543.  Chu et al. (2001) found that the X-ray spectrum of 
NGC 6543 could be adequately fit using stellar wind abundances but not 
with abundances that had been reported for the nebular material in this 
object.  Specifically, they  found that because He and N are expected to be 
significantly more enhanced in the stellar wind than in the nebular 
material  (Aller \& Czyzk 1983; Pwa, Pottasch, \& Mo 1984; Manchado \& 
Pottasch 1989; de Koter, Hubeny, Heap, \& Lanz 1996), only models 
derived from stellar wind abundances would match the observed 
spectrum.  Chu et al. (2001) emphasized, however, that the low 
temperature they derived for the X-ray emitting material
(1.7 X 10$^6$ K) is puzzling since the expected postshock temperature for 
the fast (1750 km s$^{-1}$) stellar wind of NGC 6543 is of the order of  
10$^8$K.  They also note that the location of the X-ray emitting gas 
requires that a 
significant fraction of the X-ray emitting material be nebular.  In this 
paper, 
we re-analyze the spectrum on NGC 6543 using updated abundances 
found for the nebular material in an attempt to resolve this apparent 
anomaly.

\section{Observations}

{\it Chandra} observed NGC 6543 for 46.0~ksec (ObsID 630) 
with the Advanced CCD Imaging Spectrometer (ACIS;
Garmire~\etal~1988) as the focal plane instrument.
The telescope boresight was positioned
near the center of the spectroscopy CCD array 
(ACIS-S); NGC 6543 was
imaged on the
central back-illuminated CCD (S3), which provides 
moderate
spectral resolution (E/$\Delta$E) of $\sim$4.3 at 0.5 keV and 
$\sim$9 at 1.0 keV.
Analysis of these data have already appeared in Chu 
\etal~(2001).
Data from the
observation have been reprocessed by the Chandra X-ray
Center (CXC) subsequent to the publication by Chu
\etal~(2001);  
the analysis presented here 
was performed on the reprocessed files. 
Each spectrum was extracted using the 
{\it Chandra}
Interactive Analysis of Observations (CIAO) software 
within a region
judged to contain all the X-ray flux from the 
nebula (see Fig. 1).  
We note that the current extraction (CIAOv2.2.1) reflects upgrades to the
calibration system that were made after Chu's paper 
on this system was published.
The extracted events are 
aspect-corrected, bias-subtracted, 
energy-calibrated and  
limited to grade 02346 events (ASCA system).
The background count rate 
as determined from a large,
off-source annulus region
(30- and 50-pixel radii) was negligible in 
comparison to the source count rate.

\section{Spectral Analysis}

Spectra were extracted from the entire nebula
and from three regions corresponding to the central elliptical
shell, the northern extension, and the southern extension (see Fig. 1). 
The extractions for the entire nebula and the central region excluded
counts from the central point-source.  The divisions were made
following the regions listed by Chu~\etal~(2001).

The extracted spectra were first modeled with the abundances 
used by \citet{Chu01}:
they used abundances reported by \citet{Aller83}, 
\citet{Pwa84}, 
and \citet{Manchado89} for nebular material, and abundances reported 
by \citet{deKoter96} for the stellar wind.  
For purposes of comparison, we used both the model
used by \citet{Chu01} (VRAYMOND with adopted absorption cross
sections from Balucinska-Church and McCammon 1992; hereafter RBM) as well as
a VMEKAL model with absorption cross sections adopted from Morrison
and McCammon (1983; hereafter VMM). 
Both RBM and VMM are appropriate for
optically thin thermal plasma in ionization equilibrium.
When abundance values for elements that are fit by these models 
are not available we set the abundance to Solar. 
Abundances were held fixed and only temperature, intervening column
density, and normalization were allowed as free parameters.
We are unable to reproduce the results of 
\citet{Chu01}, finding that stellar wind abundances they used modeled the
observed X-ray spectra as well as their nebular
abundances (Figs. 2 and 3). The abundances and best-fit parameters 
are listed in Tables 1
and 2. 

We then fit the observed spectra with  
nebular abundances determined from optical observations by Hyung et al. 
(2001).  
Hyung \etal`s abundances, given relative to hydrogen by number, were
converted to abundances relative to solar following Grevesse \& Sauval
1998; the converted abundances are listed in Table 1. 
We again performed two sets of fits:  RBM and VMM. 
Fits using Hyung \etal's abundances produced good results, giving a lower 
chi-squared value than both the nebular and the stellar wind abundances 
used by Chu et al. (2001).  
The temperature did not vary significantly 
over any of the models but column density decreased by 10 percent for 
fits using VMM (Fig. 4).

Since the Hyung \etal~(2001) abundances were based on 
observations taken in only one wavelength and sampled only 
at two bright regions of the nebula, they may not
be representative of the entire nebula. 
We therefore used abundances for the nebula determined by Bernard-Salas
\etal~(2003, submitted) from multiple observations in the infrared, optical,
and ultraviolet.  These provide a better sample of the nebular region and
the multiwavelength determination is more robust.  However, we note 
that Bernard-Salas is not yet published 
and these numbers are subject to small changes 
(Bernard-Salas and S.R. Pottasch, personal communication).
Again, the abundances provided by Bernard-Salas and our best-fit parameters are
listed in Tables 1 and 2.
The model fits are shown in Figure 5.
The similarity between Hyung and Bernard-Salas nebular abundances and
the improvement in fits to the data using these abundances give us 
confidence in our result.

We fit the spectra extracted from the central shell
and the northern and southern extensions using the same abundances as
those adopted for the entire nebula.  
The main result from the regional fits is that while temperature
remains constant over the emitting region, absorption appears 
to vary significantly. 
Specifically, the intervening absorption 
is greater for the southern extension
and the central shell than  
for the northern extension (Table 3).
This result is in agreement with the predictions of \citet{Miranda92} 
and with the work of \citet{Chu01}.

\section{Summary/Conclusions}

\citet{Chu01} pioneered {\it Chandra} X-ray spectral work on the planetary
nebula NGC 6543 and found that models using abundances reported for the
stellar wind better fit the observed spectrum than those using
abundances reported for the nebula.  They concluded that the X-ray
emission from this PN arose primarily from stellar wind material but emphasized
that this result was troubling when considered with the X-ray temperature derived
from their model and with the location of the X-ray emitting gas.  
In an attempt to resolve
this issue, we re-modeled the X-ray spectrum of NGC 6543 using
abundances reported by Hyung~\etal(2001) and Bernard-Salas~\etal~(Pottasch,
personal communication). 
The resulting
models fit our spectra better than either the nebular or the stellar wind
models proposed by \citet{Chu01}, suggesting that the X-ray emission
from NGC 6543 arises primarily from nebular gas.  We point out
that our result does not require that diffuse X-ray emission
from this PN originate solely from nebular material.  
Given the quality of the spectrum, a definite separation of nebular and
stellar wind abundances cannot be achieved, and in reality, 
the X-ray emitting material may be 
some mix of stellar wind and nebular material.
Our finding that observed X-ray spectral properties from NGC 6543
allow for emission predominantly from nebular material resolves
the low temperature anomaly identified by \citet{Chu01}.     

HM was supported by the NSF REU
program at SAO.  SDV was supported in part by 
NASA Grant NAG5-6711.

\clearpage
\begin{figure}
\centering
\scalebox{0.5}{\rotatebox{0}{\includegraphics{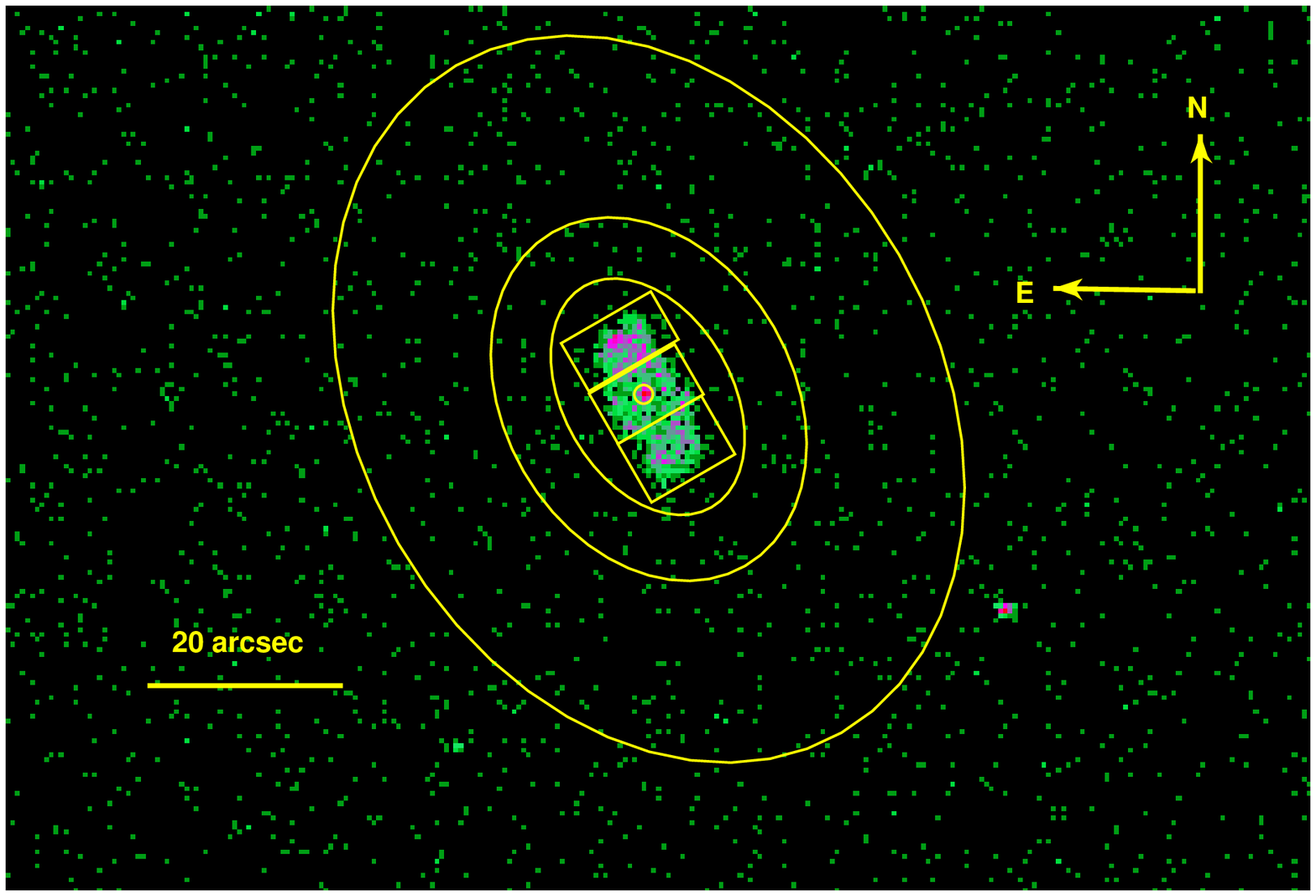}}}
\caption{X-ray image of NGC 6543 with extracted regions overlaid.
The small circle surrounds the central point source; the brightest pixel
in the center has 23 cts.
The rectangles are the northern, central, and southern regions.
The innermost oval (enclosing the rectangles) is the region 
used for the whole nebula.
The region between the two outer ovals was used to determine the background.
\label{regions_6543}}
\end{figure}

\clearpage
\begin{figure}
\centering
\scalebox{0.4}{\rotatebox{-90}{\includegraphics{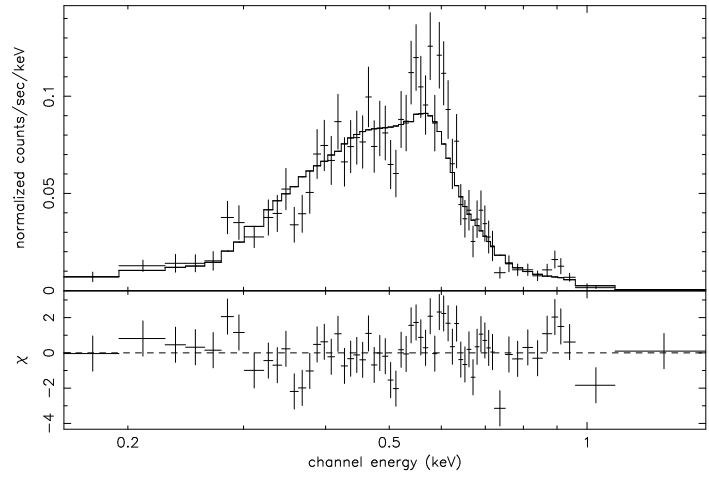}}}
\caption{X-ray spectrum of NGC 6543 with Chu stellar model
overlaid.
\label{chu_stl_6543}}
\end{figure}

\clearpage
\begin{figure}
\centering
\scalebox{0.4}{\rotatebox{-90}{\includegraphics{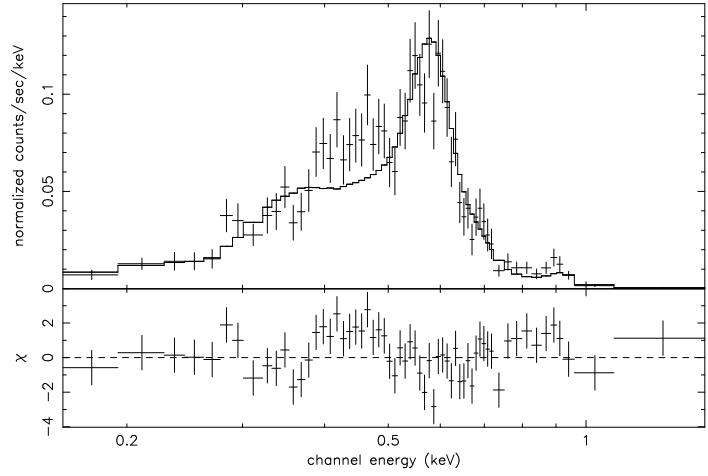}}}
\caption{X-ray spectrum of NGC 6543 with Chu nebular model
overlaid.
\label{chu_neb_6543}}
\end{figure}

\clearpage
\begin{figure}
\centering
\scalebox{0.4}{\rotatebox{-90}{\includegraphics{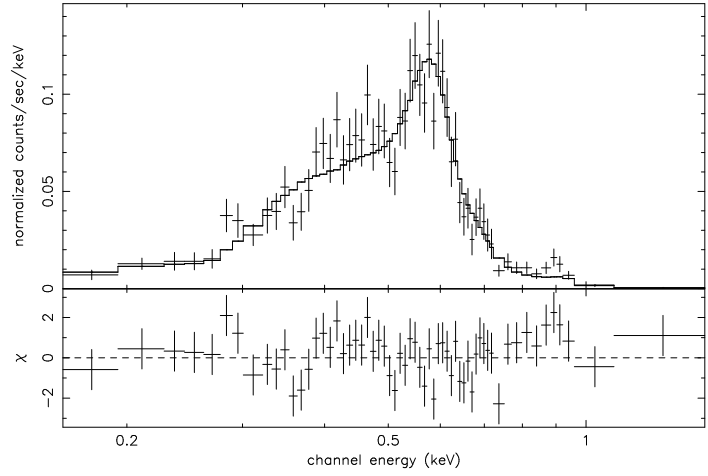}}}
\caption{X-ray spectrum of NGC 6543 with Hyung nebular model
overlaid.
\label{hyung_neb_6543}}
\end{figure}

\clearpage
\begin{figure}
\centering
\scalebox{0.4}{\rotatebox{-90}{\includegraphics{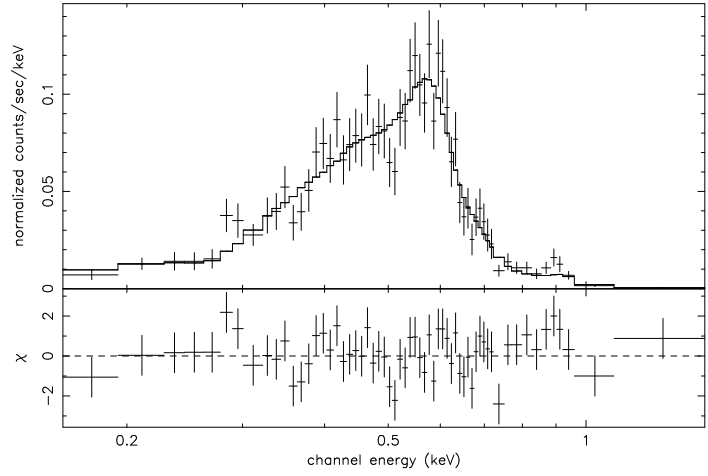}}}
\caption{X-ray spectrum of NGC 6543 with Bernard-Salas nebular model
overlaid.
\label{bs_neb_6543}}
\end{figure}

\clearpage
\begin{deluxetable}{lcccccccc}
\tabletypesize{\scriptsize}
\tablecolumns{9}
\tablewidth{0pt}
\tablecaption{Abundances (relative to Solar) used to fit NGC 6543 X-ray Spectra}
\tablehead{\colhead{Element}&\multicolumn{2}{c}{Chu Nebular\tablenotemark{1}}
&\multicolumn{2}{c}{Chu Stellar\tablenotemark{1}}&\multicolumn{2}{c}{Hyung Nebular\tablenotemark{2}}&\multicolumn{2}
{c}{Bernard-Salas Nebular\tablenotemark{3}}}
\startdata
He&\multicolumn{2}{c}{1.13}&\multicolumn{2}{c}{60}&\multicolumn{2}{c}{1.3}&\multicolumn{2}{c}{1}\\
C&\multicolumn{2}{c}{0.63}&\multicolumn{2}{c}{1}&\multicolumn{2}{c}{0.76}&\multicolumn{2}{c}{0.69$\pm0.2$}\\
N&\multicolumn{2}{c}{0.53}&\multicolumn{2}{c}{3}&\multicolumn{2}{c}{1.44}&\multicolumn{2}{c}{3.0$\pm1.1$}\\
O&\multicolumn{2}{c}{0.66}&\multicolumn{2}{c}{1}&\multicolumn{2}{c}{0.70}&\multicolumn{2}{c}{0.78$\pm0.2$}\\
Ne&\multicolumn{2}{c}{1.14}&\multicolumn{2}{c}{1}&\multicolumn{2}{c}{0.83}&\multicolumn{2}{c}{1.17$\pm0.4$}\\
Si&\multicolumn{2}{c}{1}&\multicolumn{2}{c}{1}&\multicolumn{2}{c}{0.23}&\multicolumn{2}{c}{1}\\
S&\multicolumn{2}{c}{1}&\multicolumn{2}{c}{1}&\multicolumn{2}{c}{0.37}&\multicolumn{2}{c}{0.56$\pm0.2$}\\
Ar&\multicolumn{2}{c}{1}&\multicolumn{2}{c}{1}&\multicolumn{2}{c}{1.2}&\multicolumn{2}{c}{1.55$\pm0.5$}\\
\enddata
\tablenotetext{1}{Chu~\etal~2001.}
\tablenotetext{2}{Hyung~\etal~2001.}
\tablenotetext{3}{Bernard-Salas~\etal~2003, submitted.}
\end{deluxetable}
\

\begin{deluxetable}{lcccccccc}
\tabletypesize{\scriptsize}
\tablecolumns{9}
\tablewidth{0pt}
\tablecaption{Best-fit parameters from fits to NGC 6543 Whole Nebula sans Point Source}
\startdata
\tablehead{\colhead{Parameter}&\multicolumn{2}{c}{Chu Nebular}&\multicolumn{2}{c}{Chu Stellar}&\multicolumn{2}{c}{Hyung Nebular}&\multicolumn{2}
{c}{Bernard-Salas Nebular}}\\
&RBM&VMM&RBM&VMM&RBM&VMM&RBM&VMM\\
N$_H$(cm$^{-2})$&9.8e20$\pm.007$&7.2e20$\pm.006$&8.4e20$\pm0.007$&7.8e20$\pm.007$e20&5.6$\pm0.6$e20&4.2$\pm0.6$e20&5.3$\pm0.6$e20&4.5$\pm0.5$e20\\
kT(keV)&0.12$\pm$.002&0.14$\pm$0.2&0.14$\pm$0.002&0.15$\pm$0.003&0.14$\pm.002$&0.15$\pm$0.002&0.15$\pm$0.002&0.15$\pm$0.002\\
$\chi^2_{\nu}$&1.3 &1.6&1.6&1.4&1.2&1.2&1.16&1.07\\\centering
\enddata
\end{deluxetable}
\

\begin{deluxetable}{lcccccccc}
\centering
\setlength{\tabcolsep}{0.06in}
\tabletypesize{\scriptsize}
\tablecolumns{9}
\tablewidth{0pt}
\tablecaption{Best-fit parameters from fits to NGC 6543 Selected Regions}
\startdata
\tablehead{\colhead{Parameter}&\multicolumn{2}{c}{Chu Nebular}&\multicolumn{2}{c}{Chu Stellar}&\multicolumn{2}{c}{Hyung Nebular}&\multicolumn{2}
{c}{Bernard-Salas Nebular}}\\
&RBM&VMM&RBM&VMM&RBM&VMM&RBM&VMM\\
NE N$_H$(cm$^{-2})$&6.2e20$\pm4$&4.6e20$\pm3$&5.4e20$\pm3$&4.7e20$\pm3$e20&2.6e20
$\pm3$&1.6$\pm1.5$e20&2.7$\pm2$e20&2.1$\pm2$e20\\
SE N$_H$(cm$^{-2})$&11.8e20$\pm10$&9.6e20$\pm10$&14.1e20$\pm6$&11.4e20$\pm6$e20&8.9
$\pm7$e20&6.7$\pm6$e20&7.2$\pm7$e20&5.9$\pm4$e20\\
CE N$_H$(cm$^{-2})$&16.4e20$\pm10$&12.6e20$\pm10$&11.3e20$\pm9$&13.8e20$\pm7$e20&
8.0$\pm7$e20&6.1$\pm6$e20&6.9$\pm4$e20&5.6$\pm4$e20\\
NE kT(keV)&0.13$\pm$.02&0.14$\pm$0.03&0.15$\pm$0.02&0.16$\pm$0.02&0.14$\pm.02$&0.15$\pm$0.0
2&0.15$\pm$0.02&0.16$\pm$0.02\\
SE kT(keV)&0.13$\pm$.04&0.14$\pm$0.04&0.14$\pm$0.02&0.14$\pm$0.02&0.14$\pm.03$&0.15$\pm$0.0
3&0.15$\pm$0.03&0.15$\pm$0.02\\
CE kT(keV)&0.11$\pm$.03&0.11$\pm$0.04&0.14$\pm$0.02&0.14$\pm$0.02&0.14$\pm.03$&0.14$\pm$0.0
4&0.14$\pm$0.02&0.15$\pm$0.02\\
\enddata
\end{deluxetable}

\end{document}